\documentclass[cmp,american]{svjour}
\usepackage{amsmath}
\usepackage{amssymb}

\usepackage[all]{xy}
\usepackage{babel}

\begin{document}

\title{Gerbes on orbifolds and exotic smooth $\mathbb{R}^{4}$}

\author{Torsten Asselmeyer-Maluga\inst{1} \and Jerzy Kr\'ol\inst{2}}
\institute{German Aero space center, Rutherfordstr. 2, 12489 Berlin, torsten.asselmeyer-maluga@dlr.de%
\and 
University of Silesia, ul. Uniwesytecka 4, 40-007 Katowice, iriking@wp.pl %
}
\date{Received: date / Accepted: date}
\communicated{name}
\maketitle
\begin{abstract}
By using the relation between foliations and exotic $\mathbb{R}^{4}$,
orbifold $K$-theory deformed by a gerbe can be interpreted as coming
from the change in the smoothness of $\mathbb{R}^{4}$. We give various
interpretations of integral 3-rd cohomology classes on $S^{3}$ and
discuss the difference between large and small exotic $\mathbb{R}^{4}$.
Then we show that $K$-theories deformed by gerbes of the Leray orbifold
of $S^{3}$ are in 1$\div$1 correspondence with some exotic smooth
$\mathbb{R}^{4}$'s. The equivalence can be understood in the sense
that stable isomorphisms classes of bundle gerbes on $S^{3}$ whose
codimension-1 foliations generates the foliations of the boundary
of the Akbulut cork, correspond uniquely to these exotic $\mathbb{R}^{4}$'s.
Given the orbifold $SU(2)\times SU(2)\rightrightarrows SU(2)$ where
$SU(2)$ acts on itself by conjugation, the deformations of the equivariant
$K$-theory on this orbifold by the elements of $H_{SU(2)}^{3}(SU(2),\mathbb{Z})$,
correspond to the changes of suitable exotic smooth structures on
$\mathbb{R}^{4}$.
\end{abstract}
\tableofcontents{}

\section{Introduction}

This paper presents further results in recognizing exotic small $\mathbb{R}^{4}$'s
as being relevant not only for classical GR but rather for the quantum
version of it. Moreover, non-standard $\mathbb{R}^{4}$ are also important
for other QFT's. As shown in \cite{AsselmeyerKrol2009}, the exotic
$\mathbb{R}^{4}$ can act like magnetic monopole to produce a quantization
of the electric charge. 

In our opinion, the possible modifications of physical theories are
caused by exotic rather than standard smoothness of open 4-manifolds.
In our previous paper \cite{AsselmeyerKrol2009} we represented
an exotic $\mathbb{R}^{4}$'s by 3-rd real de Rham cohomology classes
of a 3-sphere embedded in $\mathbb{R}^{4}$. Especially we observed
that though the smooth structure on $S^{3}$ is unique up to isotopy
in contrast to uncountable many non-diffeomorphic smoothings of the
topological ${\cal \mathbb{R}}^{4}$, one can detect these smoothings
by considering the 3-sphere inside the $\mathbb{R}^{4}$. Then we
have to consider other structures on the 3-sphere instead of the smoothness
structure. Here we used codimension-1 foliations, $S^{1}$-gerbes
and generalized Hitchin-Gualtieri structures.

The present paper extends the above program with a different view.
Now we are mainly focused on the role played by the integral 3-rd
cohomology classes of $S^{3}$ in recognizing exotic smoothness of
$\mathbb{R}^{4}$. We will describe many different representations
of the integral 3-rd cohomologies of $S^{3}$ with a strong physical
motivation. Among them there is a direct relation between the classifying
spaces of the foliation $B\Gamma_{1}$ and of the bundle gerbes $BPU(H)$
for some Hilbert space $H$. A more or less complete list of possible
interpretations of the 3-rd integer cohomology classes can be found
in the Appendix \ref{sec:Interpretations-integer-classes}. We'd like
to direct the readers attention to the connection between the integer
classes and $E_{8}$ bundles used in string or M-theory. In a future
paper we will come back to this point. As we conjecture in sec. \ref{sub:Casson-handle}
the difference between small and large exotic $\mathbb{R}^{4}$'s
is directly related to the difference between integer (or rational)
numbers and real numbers. Then groupoids and deformations by gerbes
as possible extensions of the integer classes $H^{3}(S^{3},\mathbb{Z})$
will be crucial for the correct recognition of \emph{large exotic}
$\mathbb{R}^{4}$'s. This conjecture about the relation between small
and large exotic $\mathbb{R}^{4}$ finishes section \ref{sec:Gerbes-and-exotic}.

Then in subsection \ref{sub:Bundle-gerbes-on S3} we turn to orbifold
constructions in the context of exotic smoothness, and show that an
exotic $\mathbb{R}^{4}$ (given by an integral cohomology class in
$H^{3}(S^{3},\mathbb{Z})$) corresponds to the deformed $K$-theory
on a certain orbifold. The deformation is performed via a gerbe on
the orbifold which is (Morita-equivalent to) the Leray orbifold of
$S^{3}$. This is described equivalently by bundle gerbes on $S^{3}$.
By considering the groupoid $SU(2)\times SU(2)\rightrightarrows SU(2)$,
where $SU(2)$ acts by conjugation on itself, we get in subsection
\ref{sub:Deformation-of K-theory} the deformation of the equivariant
K-theory of the groupoid as coinciding with the changes of smoothings
of $\mathbb{R}^{4}$. Then the deformation is performed via equivariant
3-rd cohomologies, $H_{SU(2)}^{3}(S^{3},\mathbb{Z})\simeq\mathbb{Z}$. 

The analysis of small exotic smooth $\mathbb{R}^{4}$'s via groupoids
and gerbes is interesting by itself, but our description may seem
redundant or optional because the twisted $K$-theory on the Leray
groupoid of $S^{3}$ is the twisted $K$-theory of $S^{3}$. However,
groupoids and gerbes generalize the ordinary smooth manifolds from
the point of view of $K$-theory, cohomology and geometry by introducing
singular orbifold-like structures. Although the description on a manifold
is local, the difference remains global. In this paper we show that
these global structures and generalized cohomologies characterize
\emph{small exotic} $\mathbb{R}^{4}$'s as in our Th.'s \ref{exoticTh1},
\ref{exoticTh2}, \ref{exoticTh3}. 

The possibility of describing the cycles of the deformation explicitly
is main advantage of the presented approach. It is an important step
toward building an exotic smooth function on $\mathbb{R}^{4}$ with
many physical applications. In particular, the results of this paper
seem to be highly relevant in the process of further uncovering the
meaning of exotic 4-smoothness in string topology and geometry as
well string compactifications, i.e. in the final formulation of quantum
gravity. We hope to address these important issues soon.

As we remark above, the usage of groupoids and gerbes generalize the
concept of an ordinary smooth manifold to make room for slightly singular
objects like orbifolds. In our approach it is possible to identify
these objects as part of the ordinary spacetime (like $\mathbb{R}^{4}$):
\emph{the main object is the Casson handle}. That is a hierarchical
object of wildly embedded disks having a tree-like (discrete) structure
which is continuous at the same time. In this object, all specific
properties of dimension 4 are concentrated. We will try to uncover
some of its physical properties in our future work.

\section{$S^{1}$- Gerbes on $S^{3}$ and exotic $\mathbb{R}^{4}$ \label{sec:Gerbes-and-exotic}}

In our previous paper \cite{AsselmeyerKrol2009} we uncover a relation
between an exotic (small) $\mathbb{R}^{4}$ and a cobordism class
of a codimension-1 foliation%
\footnote{In short, a foliation of a smooth manifold $M$ is an integrable subbundle
$N\subset TM$ of the tangent bundle $TM$. %
} on $S^{3}$ classified by the Godbillon-Vey class as element of the
cohomology group $H^{3}(S^{3},\mathbb{R})$. By using $S^{1}$-gerbes
it was possible to interpret the integer elements $H^{3}(S^{3},\mathbb{Z})$
as characteristic class of a $S^{1}$-gerbe over $S^{3}$.

\subsection{Exotic $\mathbb{R}^{4}$ and codimension-1 foliation}

Here we present the main line of argumentation in our previous paper
\cite{AsselmeyerKrol2009}:

\begin{enumerate}
\item In Bizacas exotic $\mathbb{R}^{4}$ one starts with the neighborhood
$N(A)$ of the Akbulut cork $A$ in the K3 surface $M$. The exotic
$\mathbb{R}^{4}$ is the interior of $N(A)$.
\item This neighborhood $N(A)$ decomposes into $A$ and a Casson handle
representing the non-trivial involution of the cork.
\item From the Casson handle we constructed a grope containing Alexanders
horned sphere.
\item Akbuluts construction gives a non-trivial involution, i.e. the double
of that construction is the identity map.
\item From the grope we get a polygon in the hyperbolic space $\mathbb{H}^{2}$.
\item This polygon defines a codimension-1 foliation of the 3-sphere inside
of the exotic $\mathbb{R}^{4}$ with an wildly embedded
2-sphere, Alexanders horned sphere (see \cite{Alex:24}). This foliation
agrees with the corresponding foliation of the homology 3-sphere $\partial A$.
This codimension-1 foliations of $\partial A$ is partly classified
by the Godbillon-Vey class lying in $H^{3}(\partial A,\mathbb{R})$
which is isomorphic to $H^{3}(S^{3},\mathbb{R})$.
\item Finally we get a relation between codimension-1
foliations of the 3-sphere lying in $\partial A$ and exotic $\mathbb{R}^{4}$.
\end{enumerate}
This relation is very strict, i.e. if we change the Casson handle
then we must change the polygon. But that changes the foliation and
vice verse. For the case of a codimension-1 foliation $\mathcal{F}$
we need an overall non-vanishing vector field or its dual, an one-form
$\omega$. This one-form defines a foliation iff it is integrable,
i.e.\[
d\omega\wedge\omega=0\]
and the leaves are the solutions of the equation $\omega=const.$
Now we define the one-forms $\theta$ as the solution of the equation\[
d\omega=-\theta\wedge\omega\]
and consider the closed 3-form\begin{equation}
\Gamma_{\mathcal{F}}=\theta\wedge d\theta\label{eq:Godbillon-Vey-class}\end{equation}
associated to the foliation $\mathcal{F}$. As discovered by Godbillon
and Vey \cite{GodVey:71}, $\Gamma_{\mathcal{F}}$ depends only on
the foliation $\mathcal{F}$ and not on the realization via $\omega,\theta$.
Thus $\Gamma_{\mathcal{F}}$, the \emph{Godbillon-Vey class}, is an
invariant of the foliation.

Now we will discuss an important equivalence relation between foliations,
cobordant foliations. Let $M_{0}$ and $M_{1}$ be two closed, oriented
$m$-manifolds with codimension-$q$ foliations. Then these foliated
manifolds are said to be \emph{foliated cobordant} if there is a compact,
oriented $(m+1)$-manifold with boundary $\partial W=M_{0}\sqcup\overline{M}_{1}$
and with a codimension-$q$ foliation transverse to the boundary inducing
the given foliation. The resulting foliated cobordism classes $\mathcal{F}\Gamma_{q}$
form an abelian group under disjoint union. The Godbillon-Vey class
$\Gamma_{\mathcal{F}}$ is also a foliated cobordism class and thus
an element of $\mathcal{F}\Gamma_{1}$. In \cite{Thu:72}, Thurston
constructed a codimension-1 foliation of the 3-sphere $S^{3}$ and
calculated the Godbillon-Vey classes, see the Appendix \ref{sec:Non-cobordant-foliationsS3}.
According to Haefliger (see Lawson \cite{Law:74} section 5), non-cobordant,
codimension-1 foliations of $S^{3}$ are classified by the elements
of $\pi_{3}(\mathcal{F}\Gamma_{1})$. Thurston constructed in the
work above a surjective homomorphism\[
\pi_{3}(\mathcal{F}\Gamma_{1})\twoheadrightarrow\mathbb{R}\]
and by results of Mather etc. (see Lawson \cite{Law:74} section 5
for an overview) the classes $\pi_{k}(\mathcal{F}\Gamma_{1})=0$ for
$k<3$ vanish. By the Hurewicz isomorphism, the surjective homomorphism
is now an element of $H^{3}(S^{3},\mathbb{R})=Hom(\pi_{3}(\mathcal{F}\Gamma_{1}),\mathbb{R})$.
Then the Godbillon-Vey class is an element of $H^{3}(S^{3},\mathbb{R})$
having values in the real numbers. Together with the results above
we obtained:\\
\emph{The exotic $\mathbb{R}^{4}$ (of Bizaca) is determined by the
codimension-1 foliations with non-vanishing Godbillon-Vey class in
$H^{3}(S^{3},\mathbb{R})$ of a 3-sphere seen as submanifold $S^{3}\subset\mathbb{R}^{4}$.}

\subsection{The reduction to integer classes and its interpretation\label{sub:The-reduction-to}}

In this subsection we will discuss the interpretation of the integer
classes in $H^{3}(S^{3},\mathbb{R})$ and the transition to abelian
gerbes. As discussed above, we have a partial classification of non-cobordant
codimension-1 foliation $\mathcal{F}$ by Godbillon-Vey classes as
elements of $H^{3}(S^{3},\mathbb{R})$ and its relation to exotic
$\mathbb{R}^{4}$'s. The Godbillon-Vey class is a real 3-form\[
\Gamma_{\mathcal{F}}=\theta\wedge d\theta\]
constructed from the one-form $\theta$. Now we will discuss the reduction
from the real classes in $H^{3}(S^{3},\mathbb{R})$ to the integer
classes in $H^{3}(S^{3},\mathbb{Z})$. First of all, 3-rd integral
cohomologies are isomorphism classes of projective, infinite dimensional,
bundles and gerbes playing a distinguish role for twisting $K$-theory
on manifolds and groupoids. This case is crucial for our following
constructions. Twisted $K$-theories and the above interpretation
are discussed briefly in the Appendix \ref{sub:Appendix-B--}.

Here we are interested in the interpretation of the integer classes
in our context of non-cobordant foliations of the 3-sphere. A more
or less complete list of possible interpretations for these classes
can be found in the Appendix \ref{sec:Interpretations-integer-classes}.
At first we remark that the cohomology class $\left[\Gamma_{\mathcal{F}}\right]$
is unchanged by a shift $\theta\to\theta+d\phi$ of the one-form $\theta$
by an exact form, i.e. we have gauge invariance in the physical sense.
Thus we can interprete the purely imaginary one-form $A=i\theta$
as a connection of a complex line bundle over $S^{3}$. Then the Godbillon-Vey
class is related to the abelian Chern-Simons form with action integral\[
S=\intop_{S^{3}}A\wedge dA=\intop_{S^{3}}\Gamma_{\mathcal{F}}\quad.\]
But that is only the tip of the iceberg. Denote by $\Gamma_{q}^{r}$
the set of germs of local $C^{r}$-diffeomorphisms of $\mathbb{R}^{q}$
forming a smooth groupoid. A codimension-$q$ Haefliger cocycle (over
an open covering $\mathcal{U}(X)=\left\{ \mathcal{O}_{i}\right\} _{i\in I}$)
of a space $X$ is an assignment: one assigns to each pair $i,j\in I$
a continuous map $\gamma_{ij}:\mathcal{O}_{i}\cap\mathcal{O}_{j}\to\Gamma_{q}^{r}$
such that \[
\gamma_{ij}(x)=\gamma_{ik}(x)\circ\gamma_{kj}(x)\]
for all $i,j,k\in I$ and $x\in\mathcal{O}_{i}\cap\mathcal{O}_{j}\cap\mathcal{O}_{k}$.
Then the setting $g_{ij}=d\gamma_{ij}$ in a neighborhood of $x\in\mathcal{O}_{i}\cap\mathcal{O}_{j}$
defines a $q$-dimensional vector bundle with transition function
$g_{ij}$, called the normal bundle of the foliation. Two Haefliger
structures $\mathcal{H}_{0},\mathcal{H}_{1}$ over $X$ are equivalent
if both are concordant (or cobordant), i.e. there is a Haefliger structure
$\mathcal{H}$ on $X\times[0,1]$, so that $\mathcal{H}_{k}=i_{k}^{*}\mathcal{H}$
with $i:X\to X\times[0,1]$, $i_{k}(x)=(x,k)$. Furthermore, it is
known that to every topological groupoid $\Gamma$ there is a classifying
space $B\Gamma$ (constructed for instance by Milnors join construction
\cite{Mil:56:2,Mil:56:3}). Then the equivalence classes of codimension-$q$
Haefliger structures of class $C^{r}$ over a manifold $M$ is given
by the set%
\footnote{Actually $[M,B\Gamma_{q}^{r}]$ has more structure than a set, i.e.
it defines a generalized cohomology theory like $[M,BG]$ defines
(complex) K theory für $G=SU$ or real K theory for $G=SO$.%
} $[M,B\Gamma_{q}^{r}]$. Then a given map $M\to B\Gamma_{q}^{r}$
determines a Haefliger structure. Now we will specialize to the (smooth)
codimension-1 case over the 3-sphere, i.e. we consider maps $S^{3}\to B\Gamma_{1}$
(setting $r=\infty$). Given a constant map $x_{0}\to B\Gamma_{1}$
with $x_{0}\in S^{3}$, i.e. a map from 0-skeleton of $S^{3}$ into
$B\Gamma_{1}$. Now we ask whether this can be extended over the other
skeleta of $S^{3}$ to get finally a map $S^{3}\to B\Gamma_{1}$.
The question can be answered by obstruction theory to state that the
elements of $H^{3}(S^{3},\pi_{3}(B\Gamma_{1}))$ label all possible
extensions. Using Thurston's surjective homomorphism we have uncountable
infinite possible extensions, i.e. all elements of $H^{3}(S^{3},\mathbb{R})$.

By using that machinery, we define a codimension-1 foliation of $S^{3}$
via a continuous function $f:S^{3}\to\mathbb{R}$ using the natural
embedding $i:\mathbb{R}\to\Gamma_{1}$ to obtain the Haefliger cocycle
$\gamma=i\circ f$. Alternatively we can also consider a function
$\tilde{f}:S^{3}\to S^{1}=U(1)$ with $f=i\cdot\log(\tilde{f})$ seen
as a section $\tilde{f}$ of some complex line bundle over $S^{3}$.
Every complex line bundle is given by a map into the classifying space
$BU(1)$, which is an Eilenberg-MacLane space%
\footnote{An Eilenberg-MacLane space $K(n,G)$ is a topological space (unique
up to homotopy) with the only non-vanishing homotopy group $\pi_{n}(K(n,G))=G$.
The group $G$ has to be abelian for $n>1$.%
} $K(2,\mathbb{Z})$. Thus, on the abstract level there is a map between
the smooth groupoid $\Gamma_{1}$ and the classifying space $BU(1)$
of complex line bundles. Then we have shown \\
\begin{theorem}There is a natural map from the smooth groupoid $\Gamma_{1}$
to the classifying space $BU(1)$ of complex line bundles. Then every
codimension-1 Haefliger structure over a 3-manifold $M$ as classified
by $[M,B\Gamma_{1}]$ is canonically mapped via a surjection to\[
[M,B(BU(1))]=[M,BK(2,\mathbb{Z})]=[M,K(3,\mathbb{Z})]=[M,BPU(H)]=H^{3}(M,\mathbb{Z})\]
The space $BU(1)$ is homotopy-equivalent to the infinite dimensional
projective space $\mathbb{C}P^{\infty}=PU(H)$ which is the Eilenberg-MacLane
space $K(2,\mathbb{Z})$ and we have $B(BU(1))=BPU(H)$ where $PU(H)$
is the projective unitary group over some seperable Hilbert space
$H$. 

The mapping above induces a mapping between $B\Gamma_{1}$ and the
corresponding classifying space $BPU(H)$ of bundle gerbes. \end{theorem} 

The close relation between codimension-1 foliations and bundle gerbes
together with the relation to (small) exotic $\mathbb{R}^{4}$ opens
a new interpretation of the integer classes $H^{3}(S^{3},\mathbb{Z})$.
In Appendix \ref{sec:Non-cobordant-foliationsS3} we will present
the construction of uncountable infinite non-cobordant codimension-1
foliations of the 3-sphere $S^{3}$. Main part in the construction
is the usage of a polygon $P$ in the hyperbolic space $\mathbb{H}$.
The volume of $P$ is proportional to the Godbillon-Vey class of the
foliation, i.e. one gets real numbers for this class. Thus, if we
restrict ourself to the integers, we will obtain integer values \[
vol(P)=n\in\mathbb{Z}\]
In the construction of the foliation, the polygon $P$ represents
some leaves. Thus if we choose an integer Godbillon-Vey class for
the foliation then these leaves have a quantized volume.

\subsection{\label{sub:Casson-handle}Small versus large exotic $\mathbb{R}^{4}$}

In this subsection we will discuss the difference between small and
large exotic $\mathbb{R}^{4}$ having omitted up to now. The non-interesting
reader can switch to the next section without loosing any substantial
material.

A small exotic $\mathbb{R}^{4}$ can be embedded smoothly into a 4-sphere
whereas a large exotic $\mathbb{R}^{4}$ cannot. Thus the construction
of both classes are rather different. As mentioned above, the small
exotic $\mathbb{R}^{4}$ can be constructed by using the failure of
the smooth h-cobordism theorem. For the large exotic $\mathbb{R}^{4}$,
one considers non-smoothable, closed 4-manifolds and constructs an
exotic $\mathbb{R}^{4}$ inside. Our result above uses extensively
Bizaca's construction of a small exotic $\mathbb{R}^{4}$ by using
the Akbulut cork for a pair of non-diffeomorphic, but homeomorphic
4-manifolds. But what can we say about large exotic $\mathbb{R}^{4}$'s?

Given a compact, simply-connected, closed 4-manifold $M$. As shown
by Freedman \cite{Fre:82}, this manifold is completely determined
by a quadratic form, the \emph{intersection form}, over the second
homology group $H_{2}(M,\mathbb{Z})$. Lateron Donaldson \cite{Don:83}
showed that not all 4-manifolds $M$ are smoothable. We don't want
to speak about the details and refer the reader to the books \cite{GomSti:1999,Asselmeyer2007}.
The criteria is simple to understand: the intersection form has to
be diagonal or must be diagonalizable%
\footnote{The diagonal values of the intersection form for a smooth 4-manifold
have a simple interpretation as self-intersections of surfaces given
by the square of the first Chern class. If that values is a non-quadratic
number like $2$ then a smooth structure don't exists.%
} over the integers $\mathbb{Z}$ then $M$ admits at least one smooth
structure. As an example we consider quadratic forms made from the
parts%
\footnote{The form $E_{8}$ is the Cartan matrix of the semi-simple Lie group
$E_{8}$.%
}\[
E_{8}=\left(\begin{array}{cccccccc}
2 & 1\\
1 & 2 & 1\\
 & 1 & 2 & 1\\
 &  & 1 & 2 & 1\\
 &  &  & 1 & 2 & 1 &  & 1\\
 &  &  &  & 1 & 2 & 1\\
 &  &  &  &  & 1 & 2\\
 &  &  &  & 1 &  &  & 2\end{array}\right)\quad,\quad H=\left(\begin{array}{cc}
0 & 1\\
1 & 0\end{array}\right)\]
The form $E_{8}$ and every sum like $E_{8}\oplus E_{8}$ is not diagonalizable
over $\mathbb{Z}$. Everything changes if we add $(1)$ or $H$. As
Freedman \cite{Fre:82} showed one can construct a manifold $M$ by
using every possible quadratic form over $\mathbb{Z}$. Thus, there
is a closed, compact, simply-connected 4-manifold $|E_{8}|$ for $E_{8}$
and $H$ corresponds to $S^{2}\times S^{2}$. But $|E_{8}|$ as well
as $|E_{8}\oplus E_{8}|=|E_{8}|\#|E_{8}|$ with the connected sum
$\#$ is not smoothable. Now we consider the form\[
E_{8}\oplus E_{8}\oplus H\oplus H\oplus H=2E_{8}\oplus3H\]
corresponding to the K3 surface, which is a smooth 4-manifold but
there is no \textbf{smooth} decomposition like\[
|2E_{8}|\#3S^{2}\times S^{2}\]
The sum $3S^{2}\times S^{2}$ represents the $3H$ part in the intersection
form. Now we consider the open manifold $X=3S^{2}\times S^{2}\setminus int(D^{4})$
together with an embedding $j:X\to K3$ in the K3 surface having a
collar, i.e. a product neighborhood $C(j)=j(\partial X)\times\mathbb{R}$
of $j(\partial X)$. The open manifold $W=3S^{2}\times S^{2}\setminus j(X)$
is homeomorphic to $int(D^{4})=\mathbb{R}^{4}$ but not diffeomorphic,
because of the non-smooth 4-manifold $|2E_{8}|\setminus D^{4}$, i.e.
there is no smoothly embedded 3-sphere! Now we remark that $X$ itself
is a Casson handle. Thus in both cases of a small and a large exotic
$\mathbb{R}^{4}$, the central object is the Casson handle. But what
is the difference in the usage of the Casson handle in both construction?

In Bizaca's construction one glued the Casson handle along a 1-handle
to the Akbulut cork and considers the interior of the resulting manifold.
Then one needs a topological disk inside of the Casson handle to get
the homeomorphism to the $\mathbb{R}^{4}$. According to the reimbedding
theorems of Freedman \cite{Fre:82} such a disk exists after 6 stages
of the Casson handle. The concrete realization of such an imbedding
by Bizaca \cite{Biz:94} gives superexponential functions for the
growth of the Casson handle. As Bizaca showed, all these handles can
be considered to be equivalent for small exotic $\mathbb{R}^{4}$.
In contrast for large exotic $\mathbb{R}^{4}$ we need the knowledge
of the whole Casson handle because the construction don't depend on
the interior of the Casson handle (which is always diffeomorphic to
the standard $\mathbb{R}^{4}$) but on the {}``boundary''. Of course
there is no real boundary of the Casson handle $CH$ but after a suitable
compactification (like Shapiro-Bing or Freudenthal) one can define
a substitute, the so-called frontier. The frontier is not a manifold
but a so-called manifold factor, i.e. the factor $W$ is not a manifold
but $W\times\mathbb{R}$ is one. For the simplest, non-trivial example
of a Casson handle, the frontier is the Whitehead continuum $Wh$,
i.e. a 3-dimensional topological space not homeomorphic to $\mathbb{R}^{3}$
but where the product $Wh\times\mathbb{R}$ is homeomorphic to $\mathbb{R}^{4}$.
Sometimes one states that $Wh$ is not simply-connected at infinity.
That property characterizes also the large exotic $\mathbb{R}^{4}$:
a smoothly embedded 3-sphere exists only {}``at infinity''. Of course
this {}``3-sphere at infity'' can be also foliated to get a class
in $H^{3}(S^{3},\mathbb{R})$ but then we will get real numbers generated
by sequences of super-exponential functions. This discussion supports
the conjecture:\\
\emph{Conjecture: Small exotic $\mathbb{R}^{4}$ are characterized
by rational numbers or by real numbers coming from at most exponential
functions. Large exotic $\mathbb{R}^{4}$ are always characterized
by real numbers generated by super-exponential functions.}

\section{Abelian gerbes deforming $K$-theory of orbifolds and exotic $\mathbb{R}^{4}$\label{sec:Abelian-gerbes-deforming}}

In this section we will get a close connection between (small) exotic
$\mathbb{R}^{4}$'s and twisted $K$-theory of orbifolds where the
twisting is induced by a gerbe. The whole subject can be presented
by using the concept of a groupoid which we will introduce now.

A groupoid ${\tt G}$ is a category where every morphism is invertible.
Let $G_{0}$ be a set of objects and $G_{1}$ the set of morphisms
of ${\tt G}$, then the structure maps of ${\tt G}$ read as:

\begin{equation}
G_{1}\,_{t}\times_{s}G_{1}\overset{m}{\rightarrow}G_{1}\overset{i}{\rightarrow}G_{1}\overset{s}{\underset{t}{\rightrightarrows}}G_{0}\overset{e}{\rightarrow}G_{1}\label{eq:def-groupoid}\end{equation}
where $m$ is the composition of the composable two morphisms (target
of the first is the source of the second), $i$ is the inversion of
an arrow, $s,\, t$ the source and target maps respectively, $e$
assigns the identity to every object. We assume that $G_{0,1}$ are
smooth manifolds and all structure maps are smooth too. We require
that the $s,\, t$ maps are submersions \cite{LupercioUribe2001},
thus $G_{1}\,_{t}\times_{s}G_{1}$ is a manifold as well. These groupoids
are called \emph{smooth} groupoids. We will present a groupoid (\ref{eq:def-groupoid})
by $G_{1}\rightrightarrows G_{0}$. In general when the source and
target maps are local homeomorphisms (diffeomorphisms), the corresponding
topological (smooth) groupoid is called an \emph{\'etale} groupoid.
A natural and important equivalence relation on groupoids is the \emph{Morita
equivalence}, see \cite{LupercioUribe2001}.

Following \cite{LupercioUribe2004}, let $G$ be a proper \'etale,
smooth groupoid ${\tt G}$. We denote the class of Morita equivalent
groupoids of ${\tt G}$ as an \emph{orbifold} $Ob$. Usually one says:
the groupoid ${\tt G}$ represents $Ob$. Given a groupoid ${\tt G}$
we define $G_{i}$ by $G_{i}=G_{1\, t}\times_{s}G_{1\, t}\times_{s}..._{t}\times_{s}G_{1}$,
$i$ times, which are sets of composable arrays of morphisms, of the
length $i$. A groupoid ${\tt G}$ is \emph{Leray} when every $G_{i},\, i=0,1...$
is diffeomorphic to a disjoint union of contractible open sets. As
was shown by Moerdijk and Pronk \cite{MoerdijkPronk1999} every orbifold
can be represented by some Leray groupoid.

Given a smooth manifold $M$ we can attach to it a natural Leray groupoid
${\cal R}\rightrightarrows{\cal U}$ representing the manifold. Let
$\{U_{\alpha}\}$ be an open cover of $M$. We take the disjoint union
${\cal U}=\underset{\alpha}{\bigsqcup}U_{\alpha}$\underbar{ }as the
set of objects $G_{0}$ and ${\cal R}=\underset{(\alpha,\beta)}{\bigsqcup}U_{\alpha}\cap U_{\beta}$,
$(\alpha,\beta)\neq(\beta,\alpha)$ as the set of morphisms. Next
let us define $s$, $t$, $e$, $i$ and $m$ maps in a groupoid as
the following natural maps:

\begin{eqnarray} \nonumber 
   s|_{U_{\alpha\beta}}:U_{\alpha\beta} \to U_{\alpha},\quad t|_{U_{\alpha\beta}}:U_{\alpha\beta}\to U_{\alpha}  \\  
 e|_{U_{\alpha}}:U_{\alpha} \to U_{\alpha},\quad i|_{U_{\alpha\beta}}:U_{\alpha\beta}\to U_{\beta\alpha}\quad,m|_{U_{\alpha\beta\gamma}}: U_{\alpha\beta\gamma}\to U_{\alpha\gamma}   \end{eqnarray}where $U_{\alpha\beta}$ is $U_{\alpha}\cap U_{\beta}$ and $U_{\alpha\beta\gamma}$
is $U_{\alpha}\cap U_{\beta}\cap U_{\gamma}$ as usual. This groupoid
is constructed directly from the open cover of a manifold and is denoted
by ${\cal M}(M,U_{\alpha})$.

\subsection{Bundle gerbes on $S^{3}$ and gerbes on groupoids\label{sub:Bundle-gerbes-on S3}}

Given a (differentiable, \'etale, proper) groupoid $G_{1}\rightrightarrows G_{0}$
we can define a gerbe on it:

\begin{definition}\label{DefGerbesonGroupid}

A \emph{gerbe }over an orbifold $G_{1}\rightrightarrows G_{0}$ (over
a groupoid representing the orbifold) is a complex line bundle $L$
over $G_{1}$ provided 

\begin{enumerate}
\item $i^{\star}L\simeq L^{-1}$
\item $\pi_{1}^{\star}L\otimes\pi_{2}^{\star}L\otimes m^{\star}i^{\star}L\overset{\theta}{\simeq}1$
\item $\theta:G_{1\, t}\times_{s}G_{1}\to U(1)$ is a 2-cocycle
\end{enumerate}
$\pi_{1,2}$ are two projections from $G_{1\, t}\times_{s}G_{1}\to G_{1}$
and $\theta$ is the trivialization of the line bundle $L$.

\end{definition}

Let us recall that a gerbe on a manifold $M$ can be defined via the
following data \cite{Hitchin1999}:

\begin{enumerate}
\item A line bundle $L_{\alpha\beta}$ on each double intersection $U_{\alpha}\cap U_{\beta}$
\item $L_{\alpha\beta}\simeq L_{\beta\alpha}^{-1}$
\item There exists a 2-cocycle $\theta_{\alpha\beta\gamma}:U_{\alpha\beta\gamma}\to U(1)$
which gives the trivialization of $L_{\alpha\beta}L_{\beta\gamma}L_{\gamma\alpha}\simeq1$
on each triple intersections.
\end{enumerate}
We see that in the case of the groupoid ${\cal M}(M,U_{\alpha})$
representing a manifold $M$ and defining the gerbe on this groupoid
as in Def. \ref{DefGerbesonGroupid}, we get exactly the gerbe on
$M$ as above \cite{LupercioUribe2001}.

We can define yet another groupoid, ${\cal G}(Y,M)$, given a manifold
$M$ and a surjective submersion $\pi:Y\to M$ %
\footnote{A surjective submersion $\pi:Y\to M$ is a map which allows local
sections, i.e. locally split. This means that for every $x\in M$
there exists an open set $U$ containing $x$, $x\in U$ and a local
section $s:U\to Y$, i.e. $s\circ\pi=id$. Locally split map is necessary
surjective \cite{MathaiMurray2001}. %
}. We need to specify the following data:

\begin{enumerate}
\item $G_{1}=Y_{\,\pi}\times_{\pi}Y=:\, Y^{[2]}$
\item $G_{0}=Y$
\item $s=p_{1}:Y^{[2]}\to Y$, $s(y_{1},y_{2})=y_{1}$, $t=p_{2}:Y^{[2]}\to Y$,
$t(y_{1},y_{2})=y_{2}$
\item $m((y_{1},y_{2}),(y_{2},y_{3}))=(y_{1},y_{3})$
\item $(y_{1},y_{2})^{-1}=(y_{2},y_{1})$
\end{enumerate}
\begin{definition}\label{enu:Def2}

A bundle gerbe over manifold $M$ is a pair $(L,Y)$ where $Y$ is
a surjective submersion and $L\overset{p}{\to}Y^{[2]}$ is a line
bundle such that

\begin{enumerate}
\item $L_{(y,y)}\simeq{\mathbf{C}}$
\item $L_{y_{1},y_{2}}\simeq L_{y_{2},y_{1}}^{\star}$
\item $L_{(y_{1},y_{2})}\otimes L_{(y_{2},y_{3})}\simeq L_{(y_{1},y_{3})}$
\end{enumerate}
\end{definition}

Now we state the following fact (Lemma 7.3.3. in \cite{LupercioUribe2001}): 

\emph{A bundle gerbe $(L,Y)$ over $M$ is the same as a gerbe over
the groupoid ${\cal G}(Y,M)$}, which is a direct consequence of Defs.
\ref{DefGerbesonGroupid} and \ref{enu:Def2}. 

Let ${\cal L}(Y,M)$ be a gerbe over the groupoid ${\cal G}(Y,M)$.
This is the bundle gerbe $(L,Y)$ on $M$. In fact bundle gerbes over
$M$ form a group with the tensor product of bundles as the group
operation \cite{LupercioUribe2001}. This group is homomorhic with
$H^{3}(M,\mathbb{Z})$ where the homomorphism is defined via the Dixmier-Duady
class $d(L,Y)$ of the bundle gerbe $(L,Y)$ \cite{LupercioUribe2001}. 

Let $Gb({\cal M}(M,U_{\alpha}))$ be the group of gerbes on the Leray
groupoid ${\cal M}(M,U_{\alpha})$ of the manifold $M$. In fact the
group of bundle gerbes on $M$ is isomorphic with the group of gerbes
on ${\cal M}(M,U_{\alpha})$ (\cite{LupercioUribe2001}, Corollary
7.3.5.).

Gerbes on groupoids are naturally related via Morita equivalence similarly
as groupoids are, where orbifolds represent their Morita equivalence
class. In fact there is a bundle gerbe representing every Morita equivalence
class of gerbes over $M$. From the other side, a natural relation
for bundle gerbes is the \emph{stable isomorphism} of these, since
they are defined via bundles. More precisely, given two bundle gerbes
$(L_{1,}Y_{1})$, $(L_{2},Y_{2})$ on $M$, we say that they are stable
isomorphic if there exist \emph{trivial} bundle gerbes on $M$ given
by bundles $T_{1}$, $T_{2}$, such that the bundles 

\begin{equation}
L_{1}\otimes T_{1}\simeq L_{2}\otimes T_{2}\end{equation}
are isomorphic. The trivial bundle gerbe is one whose Dixmier-Duady
class in $H^{3}(M,\mathbb{Z})$ is $0$, i.e $d(T_{1})=d(T_{2})=0\in H^{3}(M,\mathbb{Z})$.
It holds (\cite{LupercioUribe2001}, Corollary 7.3.10.):

\emph{Two bundle gerbes $(P,Y)$ and $(Q,Z)$ are stably isomorphic
if and only if $d(P)=d(Q)$.}

Up to the stable isomorphisms the groups of bundle gerbes on $M$
and $H^{3}(M,\mathbb{Z})$ are in fact isomorphic (\cite{LupercioUribe2001}
Theorem 7.3.13.):

\emph{There is a one-to-one correspondence between stably isomorphism
classes of bundle gerbes over $M$ and classes in $H^{3}(M,\mathbb{Z})$.
The category of bundle gerbes over $M$ with stable isomorphisms is
equivalent to the category of gerbes over $M$ with Morita equivalences. }

Thus an action of an element of $H^{3}(M,\mathbb{Z})$ can be determined
equivalently as the suitable action of the bundle gerbes whose Dixmier-Duady
class in $H^{3}(M,\mathbb{Z})$ is the element we began with. In \cite{AsselmeyerKrol2009}
we assigned non-standard smoothings of $\mathbb{R}^{4}$ to the elements
from $H^{3}(S^{3},\mathbb{Z})$ hence the action of bundle gerbes
on $S^{3}$ can be correlated with the changes of the smoothings.
In fact we are interested in twisting $K$-theories of the Leray groupoid
of $S^{3}$ by bundle gerbes on $S^{3}$. Similarly as defining the
$K$-theory for spaces and manifolds one can develop whole theory
of bundles, cohomologies and $K$-theories on the groupoids representing
orbifolds. This was performed by several authors (see e.g. \cite{LupercioUribe2001,MathaiMurray2001,AdemRuan2001,ChenRuan2000}).
In fact mathematical development of the subject was motivated by the
attempts in theoretical physics to formulate string theory on orbifolds
and the need to use twisted $K$-theoretic classes of spacetime in
order to classify the brane charges \cite{Witten1998,Witten2000}.
This is also one of the motivation for our approach to exotic smoothness
by twisted (equivariant) cohomologies: they can uncover some fundamental
relation of exotica with string theory hence QG. Besides these rather
abstract constructions are possibly relevant for the large exotic
smoothness of $\mathbb{R}^{4}$ case. Both topics we want to present
in a separate work. 

Crucial for the twisted $K$-theory are bundle gerbe modules over
$(L,Y)$. In fact given a bundle gerbe ${\cal L}(L,Y)$ on $M$ we
can define the category of bundle gerbe modules over $(L,Y)$ (see
the Appendix \ref{sub:Appendix-B--}). However, this category is equivalent
to the category of ${\cal L}(L,Y)$-twisted vector bundles over ${\cal G}(Y,M)$.
The isomorphism classes, completed by the Grothendieck procedure to
a group, gives rise to the twisted $K$-theory of the groupoid ${\cal G}(Y,M)$,
i.e. ${}^{{\cal L}}K_{gpd}({\cal G}(Y,M))$. 

Gerbes on the orbifold $G(Y,S^{3})$ are classified by $H^{3}(G(Y,S^{3}),\mathbb{Z})$
which is $H^{3}(S^{3},\mathbb{Z})$. Thus from the above and the results
of \cite{AsselmeyerKrol2009}, it follows:

\begin{theorem}\label{exoticTh1}

Given an exotic $\mathbb{R}^{4}$, $e$, corresponding to some integral
cohomology class $[e]\in H^{3}(S^{3},\mathbb{Z})$ the change of the
standard smoothing of $\mathbb{R}^{4}$ to the exotic one, $e$, determines
the deformation $\delta_{e}$ of the K-theory of the Leray groupoid
of $S^{3}$ by the bundle gerbe ${\cal L}\in[e]$, i.e.

\begin{equation}
\delta_{e}:K_{gpd}({\cal G}(Y,S^{3}))\rightarrow{}^{{\cal L}}K_{gpd}({\cal G}(Y,S^{3}))\label{eq:exoticTh1Formula}\end{equation}
 where $S^{3}$ lies at the boundary of the Akbulut
cork of $e$.

\end{theorem}

We say that the exotic structure $e$ \emph{deforms }the $K$-theory
as above. Let us see how to construct the deformation from a given
$e$. $e$ determines the codimension-1 foliation of $S^{3}$ and
its Godbillon-Vey class $[e]\in H^{3}(S^{3},\mathbb{R})$ which in
our case is integral. From the class $[e]\in H^{3}(S^{3},\mathbb{Z})$
we have the corresponding bundle gerbe ${\cal L}\in[e]$ representing
the class, i.e. $d({\cal L})=[e]$. Now the deformation of the $K$-theory
by ${\cal L}$ is well defined (see the Appendix \ref{sub:Appendix-B--})
and (\ref{eq:exoticTh1Formula}) expresses it.

We can be more explicit with the twisting of the $K$-theory of the
Leray groupoid:

The twisted $K$-theory of the Leray groupoid of $S^{3}$ is the twisted
$K$-theory of $S^{3},$ since $B(G(Y,S^{3}))=S^{3}$ and gerbes on
the orbifold $G(Y,S^{3})$ are classified by $H^{3}(G(Y,S^{3}),\mathbb{Z})$
which is $H^{3}(S^{3},\mathbb{Z})$.

Hence we can directly compute twisted cohomology ${}^{{\cal L}}K_{gpd}({\cal G}(Y,S^{3}))$
as $K^{\tau}(S^{3})$ where $\tau=[{\cal L}]\in H^{3}(S^{3},\mathbb{Z})$.
This last, following \cite{FreedHopkinsTeleman2002}, the example
1.4, reads as 

\[
K^{k}(S^{3},n[\,])=K^{\tau+k}(S^{3})=\begin{cases}
0 & ,\, k=0\\
\mathbb{Z}/n & ,\, k=1\end{cases}\]
where $\tau=n[\,]\in H^{3}(S^{3},\mathbb{Z})$ and $[\,]$ is the
generator. This twisting is given by 

\begin{equation}
\delta_{n}:K^{1}(S^{3})=\mathbb{Z}\to\mathbb{Z}/n=K^{(n)+1}(S^{3})=K^{1}(S^{3},n[\,])\end{equation}
and reflects the effect of the change of the standard smooth $\mathbb{R}^{4}$
to the exotic one, corresponding to the integral class $n[\,]\in H^{3}(S^{3},\mathbb{Z})$.
We see that the effects are detectable in generalized twisted $K$-theory.

\subsection{The deformation of the K-theory of the groupoid $SU(2)\times SU(2)\rightrightarrows SU(2)$
and exotic $\mathbb{R}^{4}$\label{sub:Deformation-of K-theory}}

Consider the hyperbolic plane $\mathbb{H}^{2}$ and its unit tangent
bundle $T_{1}\mathbb{H}^{2}$ , i.e the tangent bundle $T\mathbb{H}^{2}$
where every vector in the fiber has norm $1$. Thus the bundle $T_{1}\mathbb{H}^{2}$
is a $S^{1}$-bundle over $\mathbb{H}^{2}$. There is a foliation
$\mathcal{F}$ of $T_{1}\mathbb{H}^{2}$ invariant under the isometries
of $\mathbb{H}^{2}$ which is induced by bundle structure and by a
family of parallel geodesics on $\mathbb{H}^{2}$. The foliation $\mathcal{F}$
is transverse to the fibers of $T_{1}\mathbb{H}^{2}$. Let $P$ be
any convex polygon in $\mathbb{H}^{2}$. We will construct a foliation
$\mathcal{F}_{P}$ of the three-sphere $S^{3}$ depending on $P$.
Let the sides of $P$ be labelled $s_{1},\ldots,s_{k}$ and let the
angles have magnitudes $\alpha_{1},\ldots,\alpha_{k}$. Let $Q$ be
the closed region bounded by $P\cup P'$, where $P'$ is the reflection
of $P$ through $s_{1}$. Let $Q_{\epsilon}$, be $Q$ minus an open
$\epsilon$-disk about each vertex. If $\pi:T_{1}\mathbb{H}^{2}\to\mathbb{H}^{2}$
is the projection of the bundle $T_{1}\mathbb{H}^{2}$, then $\pi^{-1}(Q)$
is a solid torus $Q\times S^{1}$(with edges) with foliation $\mathcal{F}_{1}$
induced from $\mathcal{F}$. For each $i$, there is an unique orientation-preserving
isometry of $\mathbb{H}^{2}$, denoted $I_{i}$, which matches $s_{i}$
point-for-point with its reflected image $s'_{i}$. We glue the cylinder
$\pi^{-1}(s_{i}\cap Q_{\epsilon})$ to the cylinder $\pi^{-1}(s'_{i}\cap Q_{\epsilon})$
by the differential $dI_{i}$ for each $i>1$, to obtain a manifold
$M=(S^{2}\setminus\left\{ \mbox{\mbox{k} punctures}\right\} )\times S^{1}$,
and a (glued) foliation $\mathcal{F}_{2}$, induced from $\mathcal{F}_{1}$.
To get a complete $S^{3}$, we have to glue-in $k$ solid tori for
the $k$ $S^{1}\times\mbox{punctures}.$ Now we choose a linear foliation
of the solid torus with slope $\alpha_{k}/\pi$ (Reeb foliation).
Finally we obtain a smooth codimension-1 foliation $\mathcal{F}_{P}$
of the 3-sphere $S^{3}$ depending on the polygon $P$.

Given the conjugation classes of $SU(2)$ on $SU(2)$ (these are 2-spheres,
$S^{2}$, and 2 poles) the natural $\mathbb{Z}_{2}$- involution changes
the classes and fixes the equator $S^{2}$. As follows from \cite{AsselmeyerKrol2009}
such an involution determines the standard smooth structure on $\mathbb{R}^{4}$
whereas non-zero 3-rd integral cohomologies $H^{3}(S^{3},\mathbb{Z})$
correspond to some exotic smooth $\mathbb{R}^{4}$'s, $R_{k}^{4},k\in\mathbb{Z}$.
Now we change slightly the view and consider the involution induced
by an action of the $SU(2)$ on itself. Then we obtain elements in
the equivariant cohomology $H_{SU(2)}^{3}(SU(2),\mathbb{Z})$. By
using that idea we will get an unexpected relation to the Verlinde
algebra where the level is determined by an element in $H^{3}(S^{3},\mathbb{Z})$.
As we discussed in subsection \ref{sub:The-reduction-to} that level
can be interpreted as a surface in the hyperbolic space with quantized
volume.

Similarly as elements of $H^{3}(S^{3},\mathbb{Z})$ can twist the
ordinary $K$-theory, the elements of equivariant cohomologies $H_{SU(2)}^{3}(S^{3},\mathbb{Z})$
can be used to twist equivariant $K$-theory. The untwisted equivariant
case as above corresponds to the standard $\mathbb{R}^{4}$ (0-twist).
The twisted equivariant cohomologies by non-zero 3-rd integral cohomologies
correspond to the exotic smooth $\mathbb{R}^{4}$'s. This is because
there exists a canonical map $e:H_{G}^{*}(S^{3})\to H_{H}^{*}(S^{3})$
where $H\subset G$ is a subgroup of $G$. Taking $H=\{1\}$ we have
$H_{G}^{*}(S^{3})\to H^{*}(S^{3})$. In the case of $SU(2)$, $H_{SU(2)}^{*}(S^{3},\mathbb{Z})\simeq H^{*}(S^{3},\mathbb{Z})\simeq\mathbb{Z}$,
thus the equivariant twisting corresponds to the non-equivariant by
$e_{k}:k[\,]\to k[\,]_{eq}$ where $[\,]$ and $[\,]_{eq}$ mean the
generators of $H^{3}(S^{3},\mathbb{Z})$ and $H_{SU(2)}^{3}(S^{3},\mathbb{Z})$
correspondingly. We say that a (bundle) gerbe $d({\cal L)}=[e]\in H^{3}(S^{3},\mathbb{Z})$
\emph{twists} the equivariant $K$-theory of $SU(2)$ acting on $SU(2)$
by conjugation when the equivariant class $e_{k}([e])\in H_{SU(2)}^{3}(S^{3},\mathbb{Z})$
twists the equivariant cohomology. Again, assigning to gerbes some
non-standard small smoothings of $\mathbb{R}^{4}$, where $S^{3}\subset\mathbb{R}^{4}$,
the discussion above, the result of \cite{Freed2001,FreedHopkinsTeleman2002}
and the example 7.2.17 in \cite{LupercioUribe2001} give the following
correspondence: 

\begin{theorem}\label{exoticTh2}

Given an exotic $\mathbb{R}^{4}$, $e$, corresponding to some integral
cohomology class $[e]=k[\,]\in H^{3}(S^{3},\mathbb{Z})$, the change
of the standard smoothing of $\mathbb{R}^{4}$ to the exotic one,
$e$, determines the twisting $\delta_{e}$ of the equivariant K-theory
of the groupoid $SU(2)\times SU(2)\rightrightarrows SU(2)$ by the
gerbe ${\cal L}_{k}$ over this groupoid where $d{\cal (L}_{k})=e_{k}([e])\in H_{SU(2)}^{3}(S^{3},\mathbb{Z})$,
i.e. the twisting of the equivariant K-theory of $SU(2)$ acting on
itself by conjugation

\begin{equation}
\delta_{e}:K_{SU(2)}(SU(2))\rightarrow{}^{{\cal L}_{k}}K_{SU(2)}(SU(2))\label{eq:exoticTh2Formula}\end{equation}
 where $S^{3}\simeq SU(2)$ is the 3-sphere lying
at the boundary of the Akbulut cork of $e$ and the codimension-1
foliations of this 3-sphere generates the codimension-1 foliations
of the boundary. 

\end{theorem} Following \cite{FreedHopkinsTeleman2002} Ex. 1.7,
we can explicitly compute the ,,exotic twisting'' of the equivariant
$K$-theory:

\[
K_{SU(2)}^{n}(SU(2))=K_{SU(2)}^{(0)+n}(SU(2))\to K_{SU(2)}^{\tau+n}(SU(2))=\begin{cases}
0 & ,\, n=0\\
R(SU(2)/(\rho_{k-1}) & ,\, n=1\end{cases}\]
where $\tau=k[\,]\in H^{3}(S^{3},\mathbb{Z})$ twists the ordinary
equivariant $K$-groups, in fact $K_{SU(2)}^{0,dim(SU(2))}=K_{SU(2)}^{0,1}(SU(2))$
by Bott's periodicity, and determines exotic $\mathbb{R}^{4}$, $(\rho_{l})$
are up to $l$-dimensional representations of $SU(2)$ and ${\cal R}(SU(2))$
is the ring of the representations of $SU(2)$. 

Composing the Theorem \ref{exoticTh2}. with the result of Freed,
Hopkins and Teleman \cite{FreedHopkinsTeleman2002}, we arrive at
the following formulation: 

\begin{theorem}\label{exoticTh3}

Given two exotic $\mathbb{R}^{4}$'s, $e_{k}$, $e_{k'}$ corresponding
to the integral cohomology classes $[e_{k}]=\left(5+k\right)[\,])$
and $[e_{k'}]=\left(5+k'\right)[\,]\in H^{3}(S^{3},\mathbb{Z})$ where
$[\,]$ is the generator of $H^{3}(S^{3},\mathbb{Z})$, the change
of the smoothing of $\mathbb{R}^{4}$ from $e_{k}$ to $e_{k'}$,
determines the shift of the Verlinde algebra of $SU(2)$ from the
level $k$ to $k'$:

\begin{equation}
V_{k}(SU(2))\to V_{k'}(SU(2))\label{eq:exoticVerlindeTh3Formula}\end{equation}

\end{theorem} This is based on the relation ${\cal R}(SU(2))/(\rho_{k-1})=V_{k}(SU(2))$.
Here, one has $[e_{k}]=\left(3+2+k\right)[\,])$ where $2$ is the
Dual Coxeter number for $SU(2)$ and $dim(SU(2))=3$\cite{FreedHopkinsTeleman2002}.
It is understood that the 3-spheres lying at the
boundaries of the Akbulut corks are both the same $S^{3}=SU(2)$,
and the difference between smoothings of $\mathbb{R}^{4}$ is seen
as the shift of the levels of $V_{k}(SU(2))$ as in (\ref{eq:exoticVerlindeTh3Formula}). 

It follows that the changes between smoothings of some exotic $\mathbb{R}^{4}$'s
can be described in terms of 2-dimensional CFT or $SU(2)$ WZW models
(cf. \cite{AsselmeyerKrol2009}).

\section{Conclusion}

This paper is a natural enhancement of our previous work \cite{AsselmeyerKrol2009}.
Here we concentrated on the integer classes $H^{3}(S^{3},\mathbb{Z})$
which we interpreted as bundle gerbes. Then the full approach including
the relation to twisted $K$-theory of orbifolds was worked out to
show a relation to the Verlinde algebra. This result based on the
work in \cite{FreedHopkinsTeleman2002} is not fully unexpected. In
a ground-breaking paper of Witten \cite{Wit:89}, he related the theory
of 3-manifolds to conformal field theory by using Chern-Simons theory.
As we mentioned in subsection \ref{sub:The-reduction-to} (see also
the appendix \ref{sec:Interpretations-integer-classes} below), the
corresponding 3-form of Chern-Simons is the Godbillon-Vey invariant.
But this invariant is the key to understand exotic smoothness on 4-manifolds.
Thus we obtain a dimension ladder: a conformal field theory in 2 dimensions
determines via the level the Godbillon-Vey invariant of a codimension-1
foliation of the 3-spheres which determines the smoothness structure
on the 4-space $\mathbb{R}^{4}$ and vice versa.

The whole bunch of connections and relations in this paper are partly
related to quantum field theory. Then we may ask: Is it possible to
understand the quantization procedure in terms of exotic smoothness?
We will answer this question in the next paper by analyzing the codimension-1
foliation on the 3-sphere $S^{3}$ more carefully.

\appendix

\section*{Appendix}

\section{\label{sec:Non-cobordant-foliationsS3}Non-cobordant foliations of
$S^{3}$ detected by the Godbillon-Vey class}

In \cite{Thu:72}, Thurston constructed a foliation of the 3-sphere
$S^{3}$ depending on a polygon $P$ in the hyperbolic plane $\mathbb{H}^{2}$
so that two foliations are non-cobordant if the corresponding polygons
have different areas. We will present this construction now.

Now we consider two codimension-1 foliations $\mathcal{F}_{1},\mathcal{F}_{2}$
depending on the convex polygons $P_{1}$ and $P_{2}$ in $\mathbb{H}^{2}$.
As mentioned above, these foliations $\mathcal{F}_{1},\mathcal{F}_{2}$
are defined by two one-forms $\omega_{1}$ and $\omega_{2}$ with
$d\omega_{a}\wedge\omega_{a}=0$ and $a=0,1$. Now we define the one-forms
$\theta_{a}$ as the solution of the equation\[
d\omega_{a}=-\theta_{a}\wedge\omega_{a}\]
and consider the closed 3-form\begin{equation}
\Gamma_{\mathcal{F}_{a}}=\theta_{a}\wedge d\theta_{a}\label{eq:Godbillon-Vey-class}\end{equation}
 associated to the foliation $\mathcal{F}_{a}$. As discovered by
Godbillon and Vey \cite{GodVey:71}, $\Gamma_{\mathcal{F}}$ depends
only on the foliation $\mathcal{F}$ and not on the realization via
$\omega,\theta$. Thus $\Gamma_{\mathcal{F}}$, the \emph{Godbillon-Vey
class}, is an invariant of the foliation. Let $\mathcal{F}_{1}$ and
$\mathcal{F}_{2}$ be two cobordant foliations then $\Gamma_{\mathcal{F}_{1}}=\Gamma_{\mathcal{F}_{2}}$.
In case of the polygon-dependent foliations $\mathcal{F}_{1},\mathcal{F}_{2}$,
Thurston \cite{Thu:72} obtains\[
\Gamma_{\mathcal{F}_{a}}=vol(\pi^{-1}(Q))=4\pi\cdot Area(P_{a})\]
and thus

\begin{itemize}
\item $\mathcal{F}_{1}$ is cobordant to $\mathcal{F}_{2}$ $\Longrightarrow$$Area(P_{1})=Area(P_{2})$
\item $\mathcal{F}_{1}$ and $\mathcal{F}_{2}$ are non-cobordant $\Longleftrightarrow$$Area(P_{1})\not=Area(P_{2})$
\end{itemize}
We note that $Area(P)=(k-2)\pi-\sum_{k}\alpha_{k}$. The Godbillon-Vey
class is an element of the deRham cohomology $H^{3}(S^{3},\mathbb{R})$
which will be used later to construct a relation to gerbes. Furthermore
we remark that the classification is not complete. Thurston constructed
only a surjective homomorphism from the group of cobordism classes
of foliation of $S^{3}$ into the real numbers $\mathbb{R}$. We remark
the close connection between the Godbillon-Vey class (\ref{eq:Godbillon-Vey-class})
and the Chern-Simons form if $\theta$ can be interpreted as connection
of a suitable line bundle.

\section{\label{sec:Interpretations-integer-classes}Interpretations of the
integer classes}

Apart from the bundle gerbe interpretation, we will present different
interpretations for the reduction of $H^{3}(S^{3},\mathbb{R})$ to
the integer classes in $H^{3}(S^{3},\mathbb{Z})$:

\begin{enumerate}
\item as intersection numbers in abelian Chern-Simons theory 
\item as volume quantization of some leaves in the foliation
\item as obstruction cocyle to extend a section of a Haefliger structure
\item as parity anomaly of a $SU(2)$ gauge theory coupled to a Dirac field
over the Alexander sphere
\item as equivalence classes of loop group $\Omega E_{8}$ bundles over
the 3-sphere
\item as differential character a la Cheeger-Simons
\end{enumerate}
ad 1. The cohomology class $\left[\Gamma_{\mathcal{F}}\right]$ is
unchanged by a shift $\theta\to\theta+d\phi$ of the one-form $\theta$
by an exact form, i.e. gauge invariance in the physical sense. Thus
we can interpret the purely imaginary one-form $A=i\theta$ as a connection
of a complex line bundle over $S^{3}$. Then the Godbillon-Vey class
is the abelian Chern-Simons form with action integral\[
S=\intop_{S^{3}}A\wedge dA\quad.\]
To get any restrictions for that integral, we have to consider a 4-manifold
with boundary $S^{3}$ which by using cobordism theory always exists.
There are many models for such a 4-manifold. We start with a closed
4-manifold $M$, i.e. $\partial M=\emptyset$, and cut a 4-disk $D^{4}$
with $\partial D^{4}=S^{3}$ off. Then we obtain the desired 4-manifold
$N=M\setminus D^{4}$ with $\partial N=S^{3}$ and for the integral\[
\intop_{S^{3}=\partial N}A\wedge dA=\intop_{N}dA\wedge dA\quad.\]
Now we will follow our interpretation above, that $A$ is the connection
of a complex line bundle $L$. The first Chern class $c_{1}(L)$ of
that bundle is given by $c_{1}(L)=\frac{i}{2\pi}dA$ classifying the
complex line bundles over $N$. Then we obtain for the integral\[
S=\intop_{\partial N}A\wedge dA=-4\pi^{2}\intop_{N}c_{1}\wedge c_{1}\]
as the number of self-intersections of the 2-complex $PD(c_{1}(L))$
($PD$ Poincare dual). Thus we obtain one possible interpretation
of the integer classes $H^{3}(S^{3},\mathbb{Z})$ as self-intersections
in $N$ or as intersection between a 1-complex $PD(A)$ and $PD(dA)$
in the 3-sphere $S^{3}$. As example we can use the 4-manifold $M=\mathbb{C}P^{2}$
and construct $N=\mathbb{C}P^{2}\setminus D^{4}$. Inside of $N$
there is a $\mathbb{C}P^{1}=S^{2}\subset N$ having a canonical complex
line bundle. Then the integral $S$ is the self-intersection of $\mathbb{C}P^{1}$,
i.e. the intersection form of $\mathbb{C}P^{2}$ having one single
self-intersection with $S=1$.

ad 2. A second interpretation is given by Thurstons construction (see
Appendix A) of non-cobordant, codimension-1 foliations on $S^{3}$.
He used a polygon $P$ in the hyperbolic 2-space $\mathbb{H}$ to
construction such a foliation. In the construction of the foliation,
this polygon represents some of the leaves whereas the other are given
by the Reeb components to fill in the punctures. Then the Godbillon-Vey
invariant is proportional to the volume of the polygon $P$. Thus
we obtain that the integer classes are equivalent to a quantization
of the volumina of polygons and therefore to a quantization of the
leaves of the foliation. 

ad 3. The third interpretation used a slightly generalized version
of a foliation, the Haefliger structure. The main idea was motivated
by the observation that homotopy-theoretic properties of a foliation
are similar to a bundle. But a $G-$principal bundle over $M$ is
classified by the homotopy classes $[M,BG]$. Thus, one defines a
Haefliger structure of codimension $q$ over $M$ which is classified
by $[M,B\Gamma_{q}]$. Denote by $\Gamma_{q}^{r}$ the set of germs
of local $C^{r}$-diffeomorphisms of $\mathbb{R}^{q}$ forming a topological
groupoid. A codimension-q Haefliger cocycle over an open covering
$\mathcal{U}(X)=\left\{ \mathcal{O}_{i}\right\} _{i\in I}$ of a space
$X$ is an assignment to each pair of $i,j\in I$ of a continous map
$\gamma_{ij}:\mathcal{O}_{i}\cap\mathcal{O}_{j}\to\Gamma_{q}^{r}$
such that for all $i,j,k\in I$ \[
\gamma_{ij}(x)=\gamma_{ik}(x)\circ\gamma_{kj}(x)\]
 for all $x\in\mathcal{O}_{i}\cap\mathcal{O}_{j}\cap\mathcal{O}_{k}$.
If we set $g_{ij}=d\gamma_{ij}$ in a neighborhood of $x\in\mathcal{O}_{i}\cap\mathcal{O}_{j}$
then one defines a q-dimensional vector bundle with transition function
$g_{ij}$, the normal bundle of the foliation. Two Haefliger structures
$\mathcal{H}_{0},\mathcal{H}_{1}$ over $X$ are equivalent if both
are concordant (or cobordant), i.e. there is a Haefliger structure
$\mathcal{H}$ on $X\times[0,1]$, so that $\mathcal{H}_{k}=i_{k}^{*}\mathcal{H}$
with $i:X\to X\times[0,1]$, $i_{k}(x)=(x,k)$. Furthermore, it is
known that to every topological groupoid $\Gamma$ there is a classifying
space $B\Gamma$ (Milnors join construction \cite{Mil:56:2,Mil:56:3}).
Then the equivalence classes of codimension-q Heafliger structures
of class $C^{r}$ over a manifold $M$ is given by the set%
\footnote{Actually $[M,B\Gamma_{q}^{r}]$ has more structure than a set, i.e.
it defines a generalized cohomology theory like $[M,BG]$ defines
(complex) K theory für $G=SU$ or real K theory for $G=SO$.%
} $[M,B\Gamma_{q}^{r}]$. Then a given map $M\to B\Gamma_{q}^{r}$
determines a Haefliger structure. Now we specialize to the codimension-1
case over the 3-sphere, i.e. we consider maps $S^{3}\to B\Gamma_{1}$
(setting $r=\infty$). Given a constant map $x_{0}\to B\Gamma_{1}$
with $x_{0}\in S^{3}$, i.e. a map from 0-skeleton of $S^{3}$ into
$B\Gamma_{1}$. Now we ask whether this can extend over the other
skeleta of $S^{3}$ to get finally a map $S^{3}\to B\Gamma_{1}$.
The question can be answered by obstruction theory to state that the
elements of $H^{3}(S^{3},\pi_{3}(B\Gamma_{1}))$ are all possible
extensions. Using Thurston's surjective homomorphism we have uncountable
infinite possible extensions, i.e. all elements of $H^{3}(S^{3},\mathbb{R})$.

ad 4. Finally we consider a Dirac theory coupled to a $SU(2)$ gauge
field defined over the {}``spacetime'' $\mathbb{R}\times S^{2}$
or $[0,1]\times S^{2}$. The 2-sphere is the fixed point set of the
involution of the 3-sphere, i.e. Alexanders horned sphere. Now we
consider the parity operation $P$ acting on the $S^{2}$-coordinates.
The classical action is constructed to be parity-invariant but the
quantized theory fails to have that symmetry. The problem is the appearance
of a symmetry-braking phase\[
\Phi=\exp\left(i\frac{\pi}{2}\eta\right)\]
with the $\eta$ invariant of the Dirac operator. As stated by Yoshida
\cite{Yoshida:1985}, this invariant is related to the Chern-Simons
functional\[
CS(A)=\frac{1}{8\pi^{2}}\intop_{S^{2}\times[0,1]}tr\left(A\wedge dA+\frac{2}{3}A\wedge A\wedge A\right)\]
of the $SU(2)$ gauge field. A bundle over $S^{2}\times[0,1]$ is
determined by the transition function $S^{2}\to SU(2)$ agreeing with
the transition function of a bundle over $S^{3}$. Both $SU(2)$ bundles
are trivial and we have the gauge group $\mathcal{G}=Map(S^{3},SU(2))=\Omega^{3}SU(2)$.
The isotopy group $\pi_{0}(\mathcal{G})$ or the number of connecting
components is given by\[
\pi_{0}(\mathcal{G})=\pi_{0}(\Omega^{3}SU(2))=\pi_{3}(SU(2))=\mathbb{Z}\]
and determines the group $H^{3}(SU(2),\mathbb{Z})=H^{3}(S^{3},\mathbb{Z})$
via duality and Hurewicz isomorphism. The elements $g$ of the isotopy
group $g\in\pi_{0}(\mathcal{G})$ are global gauge transformations\[
A\to A+g^{-1}dg\]
changing the Chern-Simons functional to \[
CS(A)\to CS(A+g^{-1}dg)=CS(A)+\frac{1}{8\pi^{2}}\intop_{S^{2}\times[0,1]}(g^{-1}dg)^{3}\]
The last expression, the WZW functional, admits integer values so
that the 3-form $(g^{-1}dg)^{3}$ can be seen as element of $H^{3}(S^{3},\mathbb{Z})$
via the isomorphism $S^{3}=SU(2)$.

ad 5. Consider the semi-simple Lie group $E_{8}$ as $248$-dimensional,
smooth manifold. If we introduce a twisted product $P=M*G$ to express
that $P$ is a $G$-principal bundle over $M$ then we have the splitting
(see \cite{Boya:92} (IV.1) on page 154)\[
E_{8}=S^{3}*S^{15}*S^{23}*S^{27}*S^{35}*S^{39}*S^{47}*S^{59}\]
Thus we can immediately write down the first homotopy groups:\begin{eqnarray*}
\pi_{i}(E_{8}) & = & 0\quad i<3\\
\pi_{3}(E_{8}) & = & \mathbb{Z}\\
\pi_{k}(E_{8}) & = & 0\quad4<k<15\end{eqnarray*}
Thus the $E_{8}$ is an Eilenberg-MacLane space \[
E_{8}\sim K(\mathbb{Z},3)\quad\mbox{up to 14-skeleton}\]
Now we consider an $E_{8}$ bundle over a manifold $M$ of dimension
$\dim M<15$. Such bundles are classified by the (abelian group of)
homotopy classes $[M,BE_{8}]$. The space $BE_{8}$ is then given
by\[
BE_{8}\sim K(\mathbb{Z},4)\quad\mbox{up to 14-skeleton}\]
and thus\[
[M,BE_{8}]=[M,K(\mathbb{Z},4)]=H^{4}(M,\mathbb{Z})\quad\mbox{up to 14-skeleton.}\]
Every $E_{8}$ bundle over the 3-sphere $S^{3}$ is then classified
by $[S^{3},BE_{8}]=H^{4}(S^{3},\mathbb{Z})=0$, i.e. every $E_{8}$
bundle is trivial over any 3-manifold. But by using the path fibration
of $BE_{8}$ we get the homotopy equivalence\[
\Omega BE_{8}\sim E_{8}\]
where $\Omega BE_{8}$ is the mapping space of maps $S^{1}\to BE_{8}$.
Both functors can be interchanged to get\[
B(\Omega E_{8})\sim E_{8}\]
To proof this we consider the path fibration\[
\Omega E_{8}\to PE_{8}\to E_{8}\]
where $PE_{8}$ are all maps $[0,1]\to E_{8}$ and construct via the
functor $B$ another fibration\[
B(\Omega E_{8})\to B(PE_{8})\to BE_{8}\]
The space $B(PE_{8})$ is contractable and we obtain the desired result.
Then we can construct a bundle with structure group $\Omega E_{8}$classified
by\[
[M,B(\Omega E_{8})]=[M,E_{8}]=[M,K(\mathbb{Z},3)]=H^{3}(M,\mathbb{Z})\quad\mbox{up to 14-skeleton.}\]
Then a $\Omega E_{8}$ principal bundle over $S^{3}$ is classified
by $[S^{3},B(\Omega E_{8})]=H^{3}(S^{3},\mathbb{Z})=\mathbb{Z}$.
We remark that these classes are canonical isomorphic to classes of
$E_{8}$ bundles over the 4-sphere $S^{4}$ via the isomorphism $[S^{3},B(\Omega E_{8})]=[S^{3},\Omega BE_{8}]=[S^{4},BE_{8}]$.

ad 6. Last but not least we consider the isomorphism\[
H^{2}(M,S^{1})=H^{3}(M,\mathbb{Z})\]
induced by the exact sequence\[
1\to\mathbb{Z}\to\mathbb{R}\to\mathbb{R}/\mathbb{Z}=S^{1}\to1\]
(in the sheaf-theoretic sense) and ask for the realization of classes
in $H^{2}(M,S^{1})$. In short, Cheeger and Simons \cite{CheegerSimons1985}
studied special homomorphism from the abelian group of (integer) cycles
$Z_{k}(M,\mathbb{Z})$ into the circle group $S^{1}$. Now we consider
a 3-cycle $c\in Z_{3}(M,\mathbb{Z})$ together with a 3-form $\omega$and
define the homorphism as\[
\chi(\partial c)=\exp\left(2\pi i\intop_{c}\omega\right)\]
A concrete realization of the 3-form is given by the Chern-Simons
form.

\section{\label{sub:Appendix-B--} Twisted $K$-theory over a groupoid}

It was shown in \cite{MathaiMurray2001} that the twisted $K$-theory
of a pair $(M,[H])$, where $M$ is a manifold and $[H]$ is an integral
$\breve{C}$ech class - the curvature of a gerbe, can be obtained
from the $K$-theory of the bundle gerbes representing the Morita
equivalence class of the gerbe. The twisting of the $K$-theory contains
2 different, though related via gerbes, situations. This is the case
of $[H]$ which is a torsion class in $H^{3}(M,\mathbb{Z})$ and non-torsion
$[H]$. The first case was considered by Witten \cite{Witten1998,Witten2000}
in the context of string theory and charges of $Dp$-branes. The second
is important for us since $H^{3}(S^{3},\mathbb{Z})$ is torsionless.
Non-torsion elements twisting the $K$-theory were found to be important
in string theory as well. Both cases appeared as being uniformly described
via bundle gerbes \cite{LupercioUribe2001,MathaiMurray2001}. 

One can define the $K$-theory of bundle gerbes as the Grothendieck
group of the semi-group of bundle gerbe modules. These bundle gerbes
modules are finite dimensional. When $[H]$ is torsion in $H^{3}(M,\mathbb{Z})$,
bundle gerbe $K$-theory is isomorphic to twisted $K$-theory $K(M,[H])$.
When $[H]$ is not a torsion class one should consider the lifting
bundle gerbe associated to a projective $PU({\cal H})$-bundle (which
will be explain below) with Dixmier-Douady class $[H]$. In this case
the twisted $K$-theory is the Grothendieck group of the semi-group
of, the discussed later on, ,,$U_{{\cal K}}$-bundle gerbe modules''.
These are infinite dimensional bundle gerbe modules.

Let ${\cal L}=(L,Y)$ be a bundle gerbe over a manifold $M$, as in
Sec. \ref{sub:Bundle-gerbes-on S3}. Def. \ref{enu:Def2}., and let
$E\to Y$ be a finite rank, hermitian vector bundle. If $\phi$ is
the isomorphism of hermitian bundles $\phi:L\otimes\pi_{1}^{-1}E\to\pi_{2}^{-1}E$
it can be \emph{compatible} \emph{with the bundle} \emph{gerbe multiplication}
in the sense that the map given by: 

\[
L_{(y_{1},y_{2})}\otimes(L_{(y_{2},y_{3})}\otimes E_{y_{3}})\to L_{(y_{1},y_{2})}\otimes E_{y_{2}}\to E_{y_{1}}\]
is equal to the following map:

\[
L_{(y_{1},y_{2})}\otimes(L_{(y_{2},y_{3})}\otimes E_{y_{3}})\to L_{(y_{1},y_{2})}\otimes E_{y_{2}}\to E_{y_{1}}\,.\]
 We say that this bundle gerbe \emph{acts} on $E$. 

\begin{definition} A hermitian vector bundle $E$ over $M$ is a
\emph{bundle gerbe module} if the isomorphism $\phi:L\otimes\pi_{1}^{-1}E\to\pi_{2}^{-1}E$
is compatible with the bundle gerbe multiplication as above. 

\end{definition}

Two bundle gerbe modules are \emph{isomorphic} if they are isomorphic
as vector bundles and the isomorphism preserves the action of the
bundle gerbe. Let $Mod({\cal L})$ be the set of all isomorphism classes
of bundle gerbe modules for ${\cal L}$. The set $Mod({\cal L})$
is a semi-group \cite{MathaiMurray2001}. Recall that $d({\cal L})\in H^{3}(M,\mathbb{Z})$
is the Dixmier-Douady class of a bundle gerbe ${\cal L}=(L,Y)$ as
in Sec. \ref{sub:Bundle-gerbes-on S3}.

\begin{definition}

Given a bundle gerbe ${\cal L}=(L,Y)$ with torsion Dixmier-Douady
class, $d({\cal L})\in H^{3}(M,\mathbb{Z})$, the Grothendieck group
of the semi-group $Mod({\cal L})$, is the $K$ group of the bundle
gerbe and is denoted as $K({\cal L})$.

\end{definition}

The group $K({\cal L})$ depends only on the class $d({\cal L})\in H^{3}(M,\mathbb{Z})$
since every stable isomorphism between bundle gerbes ${\cal L}$ and
${\cal J}$ defines a canonical isomorphism $K({\cal L})\simeq K({\cal J})$.
For any class $[H]$ in $H^{3}(M,\mathbb{Z})$ we can define a bundle
gerbe ${\cal L}$ with $d({\cal L})=[H]$ and its group $K({\cal L})$.
Due to the dependence on $[H]$ and the relation of bundle gerbes
with the manifold $M$ this group is sometimes denoted by $K_{bg}(M,[H])$. 

In particular it holds \cite{MathaiMurray2001}:

\begin{enumerate}
\item If ${\cal L}=(L,Y)$ is a trivial bundle gerbe then $K_{bg}({\cal L})=K(M)$
where $K(M)$ is the untwisted $K$-theory of the manifold $M$.
\item $K_{bg}({\cal L})$ is a module over $K(M)$.
\end{enumerate}
We have made use of the point 1. above for the case $M=S^{3}$ in
Sec. \ref{sub:Deformation-of K-theory}.

As was explained in Sec. \ref{sec:Interpretations-integer-classes},
given a class $[H]\in H^{3}(M,\mathbb{Z})$ we can represent it by
a projective $PU({\cal H})$ bundle $Y$ whose class is $[H]$. Here
${\cal H}$ is some separable, possibly infinite dimensional, Hilbert
space and $U({\cal H})$ is the group of unitary operators on ${\cal H}$.
This is because the classifying space of the third cohomology group
of $M$ is the Eilenberg-Maclane space $K(\mathbb{Z},3)$. The projective
unitary group on ${\cal H}$, $PU({\cal H})=U({\cal H})/U(1)$, can
be defined. A model for $K(\mathbb{Z},3)$ is the classifying space
of $PU({\cal H})$, i.e., $K(\mathbb{Z},3)=BPU({\cal H})$. This means
that $H^{3}(M,\mathbb{Z})=[M,K(\mathbb{Z},3)]=[M,BPU({\cal H})]$,
where $[X,Y]$ denotes the homotopy classes of continuous maps from
$X$ to $Y$. Thus we obtain the realization of $H^{3}(M,\mathbb{Z})$:

\emph{Isomorphism classes of principal $PU({\cal H})$ bundles over
$M$ are parametrized by $H^{3}(M,\mathbb{Z})$.}

If $Fred$ is the space of Fredholm operators on ${\cal H}$, then,
non-twisted $K$-theory of $M$ is determined by \cite{AtiyahSegal2004}

\begin{equation}
K(M)=[M,Fred]\:.\label{eq:K(M)}\end{equation}

We can associate to a class $[H]$ (torsion or not) $PU({\cal H})$
bundle $Y$ representing the class. Let \emph{$PU({\cal H})$ }acts
by conjugations on $Fred$. We can form an associated bundle 

\[
Y(Fred)=Y\otimes_{PU({\cal H})}Fred=Y\overset{\sim}{\otimes}Fred\]
Let $[M,Y(Fred)]$ denote the space of all homotopy classes of sections
of the bundle $Y(Fred)$. Then one can define the twisted $K$-theory: 

\begin{definition}

The twisted by $[H]\in H^{3}(M,\mathbb{Z})$ $K$-theory of $M$,
i.e. $K(M,[H])$ is given by the homotopy classes $[M,Y(Fred)]$ of
the sections of $Y(Fred)$, i.e. 

\begin{equation}
K(M,[H])=[M,Y(Fred)]\label{eq:K(M,[H])}\end{equation}

\end{definition}

It holds: 

\begin{theorem} 

For a torsion class $[H]\in H^{3}(M,\mathbb{Z})$ the twisted $K$
-theory and bundle gerbe $K$-theory of ${\cal {\cal L}}=(M,L)$,
coincide, i.e.

\emph{\begin{equation}
K(M,[H])=K_{bg}(M,L)\end{equation}
}

where $d({\cal L})=[H]$. \end{theorem}

In general $[H]$ need not be torsion. One can still relate the twisted
$K$-theory of a groupoid with the classes of gerbes over groupoids,
such that in the particular case of manifolds one yields the twisted
$K$-theory of these by (bundle) gerbes over the manifolds, and the
Dixmier-Duady class of the bundle gerbe is the twisting non-torsion
3-rd integral cohomology class $[H]$ \cite{LupercioUribe2001}.

Let us recall that in the case of a smooth manifold $M$ the set of
isomorphy classes of gerbes, $Gb(M)$, is the set of the homotopy
classes $[X,BPU({\cal H})]$ for some Hilbert space ${\cal H}$. Let
$\overline{PU({\cal H})}$ denotes the groupoid $\star\times PU({\cal H})\to\star$.
Then one can prove (Proposition 6.2.5. in \cite{LupercioUribe2001}: 

\emph{For an orbifold $Ob_{{\tt G}}$ given by a groupoid ${\tt G}$
we have $Gb({\tt G})=[X,\overline{PU({\cal H})}]$ where $[X,\overline{PU({\cal H})}]$
represents the Morita equivalence classes of morphisms from ${\tt G}$
to $\overline{PU({\cal H})}$. For a manifold $M$ we obtain that
$[X,\overline{PU({\cal H})}]=H^{3}(M,\mathbb{Z})=Gb(M)$} where $X$
is the groupoid representing $M$\emph{.} In what follows we will
not distinguish between groupoid, say $\overline{PU({\cal H})}$,
and the space, say $PU({\cal H})$, when the meaning of their use
is fixed by the context. 

In the case of a non-torsion class $\alpha$ on an orbifold $Ob_{{\tt G}}$
which is represented by the morphism $\alpha:{\tt G}\to PU({\cal H})$
one should somehow deal with infinite dimensional vector spaces. Following
\cite{LupercioUribe2001} let ${\cal K}$ be the space of compact
operators of a Hilbert space ${\cal H}$. Let $U_{{\cal K}}$ be the
subgroup of $U({\cal H})$ consisting of unitary operators $I+K$
where $I$ is the identity operator and $K\in{\cal K}$. If $h\in PU({\cal H})$
and $g\in U_{{\cal K}}$ then $hgh^{-1}\in U_{{\cal K}}$ and the
semiproduct $U_{{\cal K}}\overset{\sim}{\otimes}PU({\cal H})$ is
defined. 

Now the $K$-theory for an orbifold $Ob_{{\tt G}}$ represented by
the groupoid ${\tt G}$, twisted by a gerbe ${\cal L}$ with non-torsion
class $\alpha:{\tt G}\to PU({\cal H})$ can be defined (Def. 7.2.15
in \cite{LupercioUribe2001}):

\begin{definition}

Let us consider the set of isomorphisms classes of groupoid homomorphisms
$f:{\tt G}\to U_{{\cal K}}\overset{\sim}{\otimes}PU({\cal H})$ which
are lifts of the homomorphisms $\alpha:{\tt G}\to PU({\cal H})$ such
that these are compatible with the projection $q_{2}:U_{{\cal K}}\overset{\sim}{\otimes}PU({\cal H})\to PU({\cal H})$,
i.e. $q_{2}\circ f=\alpha$. This set is the groupoid $K$-theory
twisted by the gerbe ${\cal L}$ and is denoted by $^{{\cal L}}K_{gpd}({\tt G})$. 

\end{definition}

One advantage dealing with gerbes on groupoids is that this includes
twisted equivariant $K$-theory on manifolds automatically. When the
groupoid is ${\tt G}:=SU(2)\otimes SU(2)\rightrightarrows SU(2)$
as was the case in Sec. \ref{sub:Deformation-of K-theory}., and $SU(2)$
acts on itself by conjugation, one can reformulate the remarkable
result of Freed, Hopkins and Teleman \cite{FreedHopkinsTeleman2002}
in terms of ,,twistings by gerbes'', i.e. the twisted $K$-theory
on the groupoid by the class $d({\cal L})$ from $H^{3}(S^{3},\mathbb{Z})$,
or twisted equivariant $K$-theory on $S^{3}$ by non-torsion $[H]$,
is precisely $^{{\cal L}}K_{gpd}({\tt G})$ where $d({\cal L})=(dim(SU(2)+2+k)[\,]\in H^{3}(S^{3},\mathbb{Z})$.
Moreover $^{{\cal L}}K_{gpd}({\tt G})=V_{k}(SU(2))$ is the Verlinde
algebra of $SU(2)$ at level $k$ \cite{LupercioUribe2001}.

\section*{Acknowledgment}

T.A. wants to thank C.H. Brans and H. Ros\'e for numerous discussions
over the years about the relation of exotic smoothness to physics.
J.K. thanks for the explanations and suggestions given to him by Robert
Gompf several years ago, and for the influential discussions with
Jan Sladkowski.

%\bibliographystyle{plain}
%\addcontentsline{toc}{section}{\refname}\bibliography{foliation-gerbes}

\end{document}